\documentclass[aps,prl,groupedaddress,longbibliography,twocolumn,showkeys,showpacs]{revtex4-2}
\usepackage{commath}
\usepackage{mathtools}
\usepackage{amssymb,amsfonts}
\usepackage{amsmath}
\usepackage{hyperref}
\usepackage{graphicx}
\usepackage{natbib}
\usepackage{dutchcal}
\usepackage[usenames,dvipsnames]{xcolor}
\graphicspath{{Figures/}}
\newcommand{\one}{\mbox{$1 \hspace{-1.0mm}{\bf l}$}}
\begin{document}
\title{Decoherence as detector of the Unruh effect, II}

\author{Manuel de Atocha Rodr\'iguez Fern\'andez}
  \email{mda.rzfz@gmail.com}
\affiliation{Departamento de F{\'\i}sica, CUCEI, Universidad de Guadalajara, Av. Revoluci\'on 1500, Guadalajara, CP 44430, Jalisco, M\'exico}

\author{Alexander I. Nesterov}
  \thanks{Deceased}
\affiliation{Departamento de F{\'\i}sica, CUCEI, Universidad de Guadalajara, Av. Revoluci\'on 1500, Guadalajara, CP 44430, Jalisco, M\'exico}

\author{Gennady P. Berman}
  \thanks{Deceased}
\affiliation{Theoretical Division, T-4, Los Alamos National Laboratory, Los Alamos, NM 87544, USA}

\author{C. Moreno-González}
\affiliation{Departamento de F{\'\i}sica, CUCEI, Universidad de Guadalajara, Av. Revoluci\'on 1500, Guadalajara, CP 44430, Jalisco, M\'exico}

\date{\today}
\begin{abstract}
The Unruh effect remains a central topic in quantum field theory, although its direct experimental verification continues to be challenging. Recent efforts have therefore focused on indirect detection schemes in which the Unruh effect emerges as a consequence of measuring other well-established physical phenomena. Building on our previous work, where a novel Unruh–DeWitt detector was introduced to probe decoherence in a massless scalar field, we extend the analysis to the case of an electromagnetic field. We show that the decoherence decay rates differ significantly between inertial and accelerated frames. Moreover, we find that the characteristic signature of the Unruh effect, encoded in the exponential decay of coherence, becomes observable at lower accelerations than those reported in earlier studies. \\
Throughout this work, natural units are employed: $\hbar = c = k_B = 1$.
\end{abstract}
\keywords{Non-uniform accelerated frames, quantum decoherence, quantum entanglement, Unruh effect.}

\preprint{Draft. Current version: V1}

\maketitle
\section{Introduction}

It was demonstrated by Unruh \cite{UWG}, Fulling \cite{FSA} and Davies \cite{Davies_1975} that the vacuum state is observer-dependent. An observer undergoing uniform acceleration in Minkowski spacetime will perceive the vacuum as a thermal state. The effective temperature of this thermal distribution is $T_{U} = \hslash a/(2\pi c k_{B})$, where $a$ is the observer's proper acceleration, $k_{B}$  is the Boltzmann constant, $\hslash$ is the reduced Planck constant and $c$ is the speed of light. Experimental verification of the Unruh effect presents a significant challenge. Detecting a Planck distribution with $T_{U}\approx \rm{1} \rm{K}$, would require an extremely high acceleration of approximately $a \approx 2.5 \cdot 10^{20} \, \rm{m/s^2}$.  Despite these challenges, the theoretical implications of the Unruh effect have fueled a surge of research, which continues to this day (see \cite{CLH} and references therein).

A significant consequence of the Unruh effect arises from its connection to gravity. The equivalence principle, equating inertial and gravitational mass, suggests that the vacuum excitation observed in an accelerated frame should have an analogous counterpart in a gravitational field. Indeed, the Unruh temperature, $T_{U}$, shares a similar form with the Hawking temperature, $T_{H} = \hslash g/(2\pi c k_{B})$, associated with a black hole \cite{HS1}, with $g = c^{4}/(4GM)$, where $G$ is the Newton gravitational constant and $M$ is the mass of the black hole. This resemblance leads to the combined term "Hawking-Unruh temperature." Therefore, the concept of the Minkowski vacuum is modified in both accelerated frames and gravitational fields. The definition of a "particle" becomes observer-dependent, contingent on the observer's motion and the presence of gravity (\cite{BNDP, FSA1}).

A central question is how to experimentally detect these excited particles and verify their Planckian statistics. Direct observation requires exceptionally high accelerations, achievable, for example, near strong gravitational fields, like those surrounding black holes. Consequently, researchers have proposed various indirect detection methods for the Unruh effect (e.g., \cite{FHY}, \cite{UWG}, \cite{U95}, \cite{HDMM}, \cite{ZWPR}). Of particular interest are detectors based on Berry phase measurements, as suggested in \cite{MMED} and \cite{MMFE}. According to \cite{MMFE}, this proposal could significantly reduce the required acceleration, down to $a \approx 2.5 \cdot 10^{17} \, \rm m/s^2$.

In our previous work \cite{AINGPBMDAXW_DADOUDWF_2020}, we demonstrated that information about the Unruh effect is encoded in the decoherence of a detector's reduced density matrix (the exponential decay of the off-diagonal matrix elements). We showed that this phase decay could be observable at accelerations as low as $10^{13} \, \rm m/s^2$ for a massless scalar field. Such accelerations should be sustained only for a time $\tau \approx 100 \, \mu s$. This represents a significant improvement of four orders of magnitude compared to the results in \cite{MMED}, \cite{MMFE}. However, a practical challenge arises in interpreting measurements of a \textit{scalar field}. To address this, we extend our analysis to an \textit{electromagnetic field} in this work, providing a more tangible experimental framework.
\section{The decoherence function for an electromagnetic field}

We assume that the observer moves with an uniform acceleration, $a$, in the z-direction with respect to the inertial reference frame in the Minkowski spacetime. The transformations of coordinates,
\begin{align}
t = \frac{1}{a} e^{a\zeta} \sinh(a\tau), \quad
z = \frac{1}{a} e^{a\zeta} \cosh(a\tau),
\label{EQ01}
\end{align}
describe the right wedge of the Rindler spacetime whose line element is
\begin{align}
ds^2 = e^{2a\zeta} (d\tau^2 - d\zeta^2) - d\textbf{x}_{\bot}^2.
\label{EQ02}
\end{align}
Now, the actual measurements are made by our model detector \cite{AINGPBMDAXW_DADOUDWF_2020}. The detector is considered as a box containing a non-relativistic particle interacting with an electromagnetic field. It is assumed that the detector's original state is the ground state. The quanta of the electromagnetic field are registered if the detector is found in an excited state. The entire system, "detector + electromagnetic field", is governed by the Hamiltonian $\mathcal{H}_{Tot} = H_d +H_f + H_{int}$, where $H_d$ is the Hamiltonian of the detector, $H_f$ is the  Hamiltonian of the electromagnetic field, and $H_{int}$ stands for the interaction Hamiltonian. The latter can be written as $H_{int} = \int_{\Sigma_{\tau}} \mathcal{H}_{int} \,d^3 \mathbf{r}$, where $\mathcal{H}_{int}$ is the interaction Hamiltonian density, and the integral is taken over the three-dimensional Rindler surface $\Sigma_{\tau}$ at time $\tau = \rm{constant}$. In the original formulation of the Unruh-DeWitt model, the detector was considered as a pointlike particle, with the interaction Hamiltonian being $H_{int}= \int_{\Sigma_t} \delta^3 (\mathbf{r} - \mathbf{r}(\tau)) \mathcal{H}_{int} \,d^3 \mathbf{r}$, where $\mathbf{r}(\tau)$ describes the trajectory of the detector in its proper spacetime. In our present work, we consider both cases: a pointlike detector and a detector of finite size.

In our analysis, we consider a model detector as a two-level system with a transition energy $\varepsilon$. The Hamiltonian of the detector under our consideration has the form $H_d = (\varepsilon/2)\sigma_z$, where $\sigma_z$ is the Pauli matrix. We assume that, in the uniformly accelerated reference frame, the  detector is located at the origin of the coordinates. The interaction Hamiltonian $H_{int}$ can be written as:
\begin{align}
H_{int}(\Omega, \textbf{k}_{\bot}) = \int \sqrt{-g} f(\mathbf{r}) \mathcal{H}_{int}(\mathbf{r}, \tau) \, d^3 \mathbf{r},
\label{EQ03}
\end{align}
where $\mathcal{H}_{int}$ is the interaction Hamiltonian density, the integral is taken over the three-dimensional surface, $\Sigma_{\tau}$, at time $\tau = \rm{constant}$, $(\Omega, \textbf{k}_{\bot})$ are the Rindler momenta, $\mathbf{r} = (\textbf{x}_{\bot}, \zeta)$ are the Rindler coordinates and the function $f(\mathbf{r})$ describes the spatial profile of the detector in Rindler coordinates. The interaction Hamiltonian density is (\cite{BKT}):
\begin{widetext}
\begin{align}
\mathcal{H}_{int}(\mathbf{r}, \tau) = -\mu_{\zeta} \otimes B_{\zeta} = \mu_{B} \sigma_{z} \otimes \Big( i \sum_{\Omega \textbf{k}_{\bot}, \lambda} \sqrt{\frac{1}{2\Omega V}} \Big( v^{R}(\mathbf{r}, \tau) \hat{b}_{\Omega \textbf{k}_{\bot}, \lambda} - v^{R \ast}(\mathbf{r}, \tau) \hat{b}_{\Omega \textbf{k}_{\bot}, \lambda}^\dagger \big) [\hat{\mathbf{k}} \times \varepsilon_{\Omega \textbf{k}_{\bot}, \lambda}]_{\zeta} \Big),
\label{EQ04}
\end{align}
\end{widetext}
where $\mu_{\zeta}$ is the magnetic moment in the Rindler coordinates, $B_{\zeta}$ is the Rindler $\zeta$-component of the magnetic field, $\mu_{B}$ is the Bohr magneton, $\hat{b}_{\Omega \textbf{k}_{\bot}, \lambda}$ and $\hat{b}_{\Omega \textbf{k}_{\bot}, \lambda}^{\dagger}$ are the corresponding Rindler annihilation and creation operators for the electromagnetic field, $v^{R}(\mathbf{r})$ is the Rindler spacelike mode, $V$ is the volume bounded by the three-dimensional surface $\Sigma_{\tau}$, and the $\varepsilon_{\Omega \textbf{k}_{\bot}, \lambda}$ are the polarization vectors with $\lambda = 1, 2$. For Rindler coordinates, $\sqrt{-g} = e^{a\zeta}$. Making the following definitions for the form factor $g_{\Omega \textbf{k}_{\bot}, \lambda}$ (\cite{AINGPBMDAXW_DADOUDWF_2020, BKT}):
\begin{align}
g_{\Omega \textbf{k}_{\bot}, \lambda} = \int  d^3 \mathbf{r} \ e^{a\zeta} f(\mathbf{r}) v^{R}(\mathbf{r}),
\label{EQ05}
\end{align}
\begin{align}
g_{\Omega \textbf{k}_{\bot}, \lambda} = g_{\Omega \textbf{k}_{\bot}} \big[ \hat{\mathbf{k}} \times \varepsilon_{\Omega \textbf{k}_{\bot}, \lambda} \big]_{\zeta},
\label{EQ06}
\end{align}
\begin{align}
f(\mathbf{r}) = \frac{1}{(2\pi)^3} \sum_{\lambda} \int d\Omega d^{2} \textbf{k}_{\bot} e^{-a\zeta} g_{\Omega \textbf{k}_{\bot}, \lambda} v^{R *}(\mathbf{r}),
\label{EQ07}
\end{align}
\begin{align}
v^{R}(\mathbf{r}) = \sqrt{\frac{2\Omega \sinh(\pi \Omega/a)}{\pi^2 a}} e^{i\textbf{k}_{\bot} \cdot x_{\bot}} K_{\frac{i\Omega}{a}} \Big( \frac{k_{\bot}}{a} e^{a\zeta} \Big) \varepsilon_{\Omega \textbf{k}_{\bot}, \lambda},
\label{EQ08}
\end{align}
we will obtain the expression
\begin{align}
& H_{int}(\Omega, \textbf{k}_{\bot}) =  \nonumber \\
& \frac{i\Lambda_{em}}{\sqrt{V}} \sigma_z \otimes \sum_{\Omega \textbf{k}_{\bot}, \lambda} \frac{1}{\sqrt{2 \Omega}} \big( g_{\Omega \textbf{k}_{\bot}, \lambda} \hat{b}_{\Omega \textbf{k}_{\bot}, \lambda}^\dagger - g_{\Omega \textbf{k}_{\bot}, \lambda}^\ast \hat{b}_{\Omega \textbf{k}_{\bot}, \lambda} \big),
\label{EQ09}
\end{align}
where $\Lambda_{em}$ is the electromagnetic coupling constant. $\Lambda_{em} = \sqrt{\frac{\mu_{B}^{2}}{\epsilon_0 \hslash c^{3}}}$ in SI and $\Lambda_{em} = 2\sqrt{\pi} \mu_{B}$ in natural units. \\
In Eq. \ref{EQ09}, we assume that the interaction term is much smaller than the effective Zeeman interaction (\cite{AINGPBMDAXW_DADOUDWF_2020}). It is well-known that in this case only the $\sigma_{z}$ operator can be used in the interaction Hamiltonian (\cite{AINGPBMDAXW_DADOUDWF_2020}). We call the system with the Hamiltonian in Eq. \ref{EQ09} energy conserving because the operator $\sigma_{z}$ commutes with the total Hamiltonian. As a result, the initial probabilities of population of the detector do not change in time (\cite{AINGPBMDAXW_DADOUDWF_2020}). We assume that the density operator $\varrho(\tau)$ accounts for the entire system evolution in time. The computation of the decoherence function yields (see the Supplemental Material):
\begin{align}
\gamma_{0}(\tau) = \frac{\Lambda_{em}^{2}}{4 \pi^3} \int_0^\infty \frac{\abs{g_{\Omega}}^2}{\Omega^{3}} \big( 1 - \cos{(\Omega \tau)} \big) \coth \Big( \frac{\pi \Omega}{a} \Big) d\Omega.
\label{EQ10}
\end{align}
Knowing that the form factor for the pointlike detector is (with $g_{\Omega \textbf{k}_{\bot}}$ as the transverse component) (\cite{AINGPBMDAXW_DADOUDWF_2020})
\begin{align}
g_{\Omega \textbf{k}_{\bot}, \lambda} = \sqrt{\frac{4\Omega\sinh(\pi \Omega/a)}{\pi a}} K_{\frac{i\Omega}{a}} \Big( \frac{|\mathbf{k}_{\bot}|}{a} \Big) \big[ \hat{\mathbf{k}} \times \varepsilon_{\Omega \textbf{k}_{\bot}, \lambda} \big]_{\zeta},
\label{EQ11}
\end{align}
then:
\begin{align}
& |g_{\Omega}|^2 = \int \Big|\sum_{\lambda} g_{\Omega \textbf{k}_{\bot}, \lambda}\Big|^2 d^2 {\mathbf{k}}_{\perp} = \nonumber \\
& \int |\mathbf{k}_{\perp}|^{2} | g_{\Omega \textbf{k}_{\bot}}|^2 d^2 {\mathbf{k}}_{\perp} = \Big( \frac{64 \pi^2}{3} \Big) \Omega^{2} (\Omega^{2} + a^{2}).
\label{EQ12}
\end{align}
Hence:
\begin{align}
\gamma_{0}(\tau) = \frac{16 \Lambda_{em}^{2}}{3 \pi} \int_{0}^{\infty} \Big( \frac{\Omega^{2} + a^{2}}{\Omega} \Big) \big( 1 - \cos{(\Omega \tau)} \big) \coth \Big( \frac{\pi \Omega}{a} \Big) d\Omega.
\label{EQ13}
\end{align}
In order to avoid the ultraviolet divergence evidently affecting $\gamma_{0}(\tau)$ (see \cite{AINGPBMDAXW_DADOUDWF_2020}), we consider the form factor as
\begin{align}
& g_{\Omega \textbf{k}_{\bot}, \lambda} = \nonumber \\
& e^{-\frac{l \Omega}{2}} \sqrt{\frac{4 \Omega\sinh(\pi \Omega/a)}{\pi a}} K_{\frac{i\Omega}{\pi a}} \Big( \frac{|\mathbf{k}_{\bot}|}{a} \Big) \big[ \hat{\mathbf{k}} \times \epsilon_{\mathbf{k} \lambda} \big]_{\zeta},
\label{EQ14}
\end{align}
where $l$ is the characteristic size of the detector. Then, the decoherence function becomes
\begin{align}
& \gamma(\tau) = \frac{16 \Lambda_{em}^{2}}{3 \pi} \times \nonumber \\
& \int_{0}^{\infty} \Big( \frac{\Omega^{2} + a^{2}}{\Omega} \Big) e^{-l \Omega} \big( 1 - \cos{(\Omega \tau)} \big) \coth \Big( \frac{\pi \Omega}{a} \Big) d\Omega.
\label{EQ15}
\end{align}
Performing the integration we obtain \cite{HZ}:
\begin{widetext}
\begin{align}
\gamma(\tau) = & \ \frac{4 \Lambda_{em}^2}{3 \pi l^2} \Re \ln (1 + i \tau / l) - \frac{8 \Lambda_{em}^2}{3 \pi l^2} \Re \ln \bigg( \frac{\Gamma (1 + a l / 2\pi + i a\tau / 2\pi)}{\Gamma (1 + a l / 2\pi)} \bigg) + \nonumber \\
& \frac{4 \Lambda_{em}^2}{3 \pi a^2 l^4} \frac{1 + 3 l^2 / \tau^2}{(1 +  l^2 / \tau^2)^{2}} + \frac{2 \Lambda_{em}^2}{3 \pi^3 l^2} \psi_{1} (1 + al/2\pi) - \frac{2 \Lambda_{em}^2}{3 \pi^3 l^2} \psi_{1} (1 + a l / 2\pi + i a\tau / 2\pi).
\label{EQ16}
\end{align}
\end{widetext}
Here $\Gamma(z)$ is the gamma function and $\psi_{1}(z)$ is the polygamma function of first order.
Considering physical units, we find that, in the limit of $a\tau/2\pi c \gg 1$, we can approximate the decoherence function as:
\begin{align}
\gamma(\tau) \approx \frac{4\Lambda_{em}^2}{3\pi l^2} \ln (c\tau / l) + \frac{2 \Lambda_{em}^2 a \tau}{3\pi c l^2} + \frac{4 \Lambda_{em}^2 c^2}{3\pi a^2 l^4} - \frac{\Lambda_{em}^2 a\tau}{6\pi^3 c l^2}.
\label{EQ17}
\end{align}

\begin{figure}
\begin{center}
\scalebox{0.75}{\includegraphics{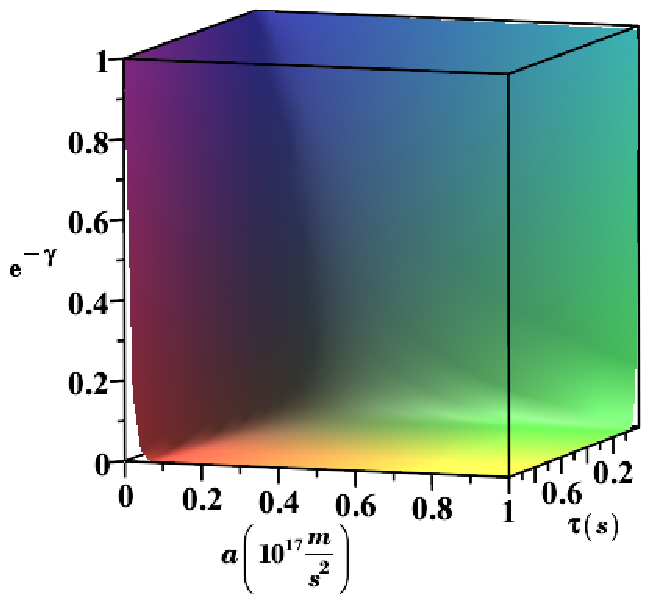}}
\end{center}
\caption{The exponential decay function for the model detector and the electromagnetic field, $e^{-\gamma(\tau)}$, versus $\tau$ and $a$.
\label{Fig1}}
\end{figure}

\begin{figure}
\begin{center}
\scalebox{0.50}{\includegraphics{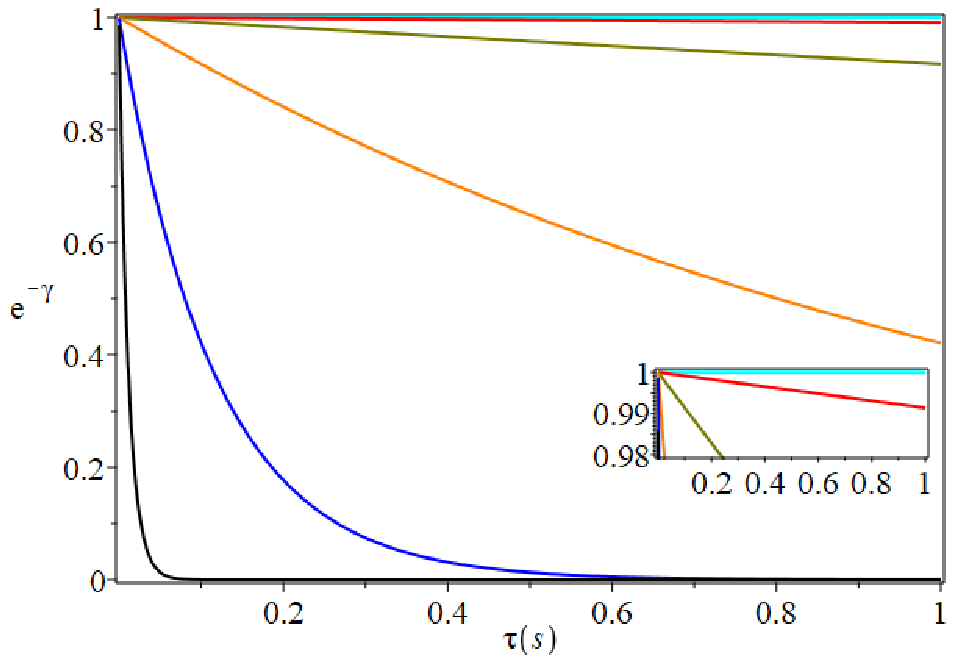}}
\end{center}
\caption{The exponential decay function for our model detector and the electromagnetic field, $e^{-\gamma(\tau)}$, versus $\tau$. From top to bottom: $a = 0$ (cyan), $a = 1 \cdot 10^{13} \ \textrm{m/s}^2$ (red), $a = 1 \cdot 10^{14} \ \textrm{m/s}^2$ (green), $a = 1 \cdot 10^{15} \ \textrm{m/s}^2$ (orange), $a = 1 \cdot 10^{16} \ \textrm{m/s}^2$ (blue), $a = 1 \cdot 10^{17} \ \textrm{m/s}^2$ (black). Inset: zoom of the main figure.
\label{Fig2}}
\end{figure}

\noindent In SI units, the electromagnetic field coupling constant has the numerical value of
\begin{align}
\Lambda_{em} = \sqrt{\frac{\mu_B^2}{\epsilon_0 \hslash c^3}} \approx 5.847 \cdot 10^{-14} 
\rm{m},
\label{EQ3080}
\end{align}
or, in the Natural units we use in this work,
\begin{align}
\Lambda_{em} = 2\sqrt{\pi} \mu_{B} \approx 2.963 \cdot 10^{-7} (eV)^{-1}.
\label{EQ3081}
\end{align}
As asserted in our previous work \cite{AINGPBMDAXW_DADOUDWF_2020}, the detector exhibits full phase decoherence, even for an inertial motion. The decoherence effect demonstrates insensitivity to the choice of the cutoff parameter, while exhibiting a strong dependence on the coupling constant, $\Lambda_{em}$, as evidenced by Fig. \ref{Fig1} and Fig. \ref{Fig2}. Comparing this result to the corresponding outcome for a scalar field (with $\Lambda_{ms}^2=10^{-6}$) obtained using our model detector in our previous work \cite{AINGPBMDAXW_DADOUDWF_2020}, we find that for accelerations on the order of $10^{12} \ \textrm{m/s}^2 \sim 10^{13} \ \textrm{m/s}^2$ (see Fig. \ref{Fig2}) the detector is sensitive to changes in the quantum decoherence induced by the Unruh effect in the presence of an electromagnetic field.
\section{Conclusions}
In this work, we have demonstrated that the Unruh effect, through its influence on the quantum vacuum, induces a measurable decay in the decoherence of a model Unruh-DeWitt detector coupled to a real electromagnetic field. This decay, observable at accelerations on the order of $10^{12} \ \textrm{m/s}^2 \sim 10^{13} \ \textrm{m/s}^2$, provides a compelling signature for the detection of this elusive effect. \\
Our analysis, building upon our previous work \cite{AINGPBMDAXW_DADOUDWF_2020} where decoherence served as an indicator of the Unruh effect for a scalar field, extends the investigation to a more experimentally viable scenario involving electromagnetic fields. The observed sensitivity of the decoherence to the coupling constant, $\Lambda_{em}$, as evidenced by the variations illustrated in Fig. \ref{Fig1} and Fig. \ref{Fig2}, highlights the potential tunability of the detector's response. \\
Future research will focus on further enhancing the sensitivity of this detection scheme, exploring alternative coupling mechanisms and detector designs. Moreover, the implications of the Unruh effect and the associated quantum entanglement on the fundamental structure of reference frames and their corresponding quantum states will be investigated. This promises to shed light on the deep connections between quantum information, gravity, and the very fabric of spacetime.
\section*{Acknowledgments}
One of the co-authors, Prof. Alexander I. Nesterov, contributed substantially to the development and revision of this work, but passed away prior to its final completion. \\
Another co-author, Prof. Gennady P. Berman, passed away before the completion of this manuscript. His contributions to the ideas and conceptual development of this work are gratefully acknowledged.
\bibliographystyle{unsrt}
\bibliography{AM}

@article{AINGPBMDAXW_DADOUDWF_2020,
title = "{Decoherence as detector of the Unruh effect}",
author = "{A. I. Nesterov, G. P. Berman, Manuel de Atocha Rodríguez-Fernández, X. Wang}",
journal = {Phys. Rev. Research},
volume = {2},
issue = {4},
pages = {043230},
numpages = {5},
year = {2020},
month = {Nov},
publisher = {American Physical Society},
doi = {10.1103/PhysRevResearch.2.043230},
url = {https://link.aps.org/doi/10.1103/PhysRevResearch.2.043230},
}

@article{HDMM,
title = "{Renormalized Unruh-DeWitt particle detector models for boson and fermion fields}",
author = "{D. H\"ummer, E. Mart\'{\i}n-Mart\'{\i}nez and A. Kempf}",
journal = {Phys. Rev. D},
volume = {93},
issue = {2},
pages = {024019},
numpages = {50},
year = {2016},
month = {Jan},
publisher = {American Physical Society},
doi = {10.1103/PhysRevD.93.024019},
url = {https://link.aps.org/doi/10.1103/PhysRevD.93.024019},
}

@article{MMED,
title = "{Berry phase as a quantum thermometer}",
author = "{E. Mart\'in-Mart\'inez, A. Dragan, R. B. Mann and I. Fuentes}",
journal = {New Journal of Physics},
volume = {15},
number = {5},
pages = {053036},
year = {2013},
month = {May},
publisher = {IOP Publishing},
doi = {10.1088/1367-2630/15/5/053036},
url = {https://doi.org/10.1088/1367-2630/15/5/053036},
}

@article{MMFE,
title = "{Using Berry's phase to detect the Unruh effect at lower accelerations}",
author = "{E. Mart\'in-Mart\'inez, I. Fuentes and R. B. Mann}",
journal = {Phys. Rev. Lett.},
volume = {107},
issue = {13},
pages = {131301},
numpages = {5},
year = {2011},
month = {Sep},
publisher = {American Physical Society},
doi = {10.1103/PhysRevLett.107.131301},
url = {https://link.aps.org/doi/10.1103/PhysRevLett.107.131301},
}

@article{FHY,
title = "{Number-conserving coherent state in Rindler space}",
author = "{F. Hong-Yi}",
journal = {Communications in Theoretical Physics},
volume = {54},
number = {3},
pages = {457},
year = {2010},
month = {Sep},
publisher = {IOP Publishing},
doi = {10.1088/0253-6102/54/3/15},
url = {https://doi.org/10.1088/0253-6102/54/3/15},
}

@Book{HZ,
title = "{{\rm eds.} NIST Handbook of Mathematical Functions}",
author = "{F. W. J. Olver,  D. W. Lozier, R. F. Boisvert and C. W. Clark}",
publisher = {Cambridge University Press, New York},
year = {2010},
}

@Book{BKT,
title = "{Perturbation Theory for Solid-state Quantum Computation With Many Quantum Bits}",
author = "{G. P. Berman, D. I. Kamenev and V. I. Tsifrinovich}",
publisher = {Rinton Pr Inc.},
year = {2005},
}

@article{CLH,
title = "{The Unruh effect and its applications}",
author = "{L. C. B. Crispino, A. Higuchi and G. E. A. Matsas}",
journal = {Rev. Mod. Phys.},
volume = {80},
issue = {3},
pages = {787--838},
numpages = {0},
year = {2008},
month = {July},
publisher = {American Physical Society},
doi = {10.1103/RevModPhys.80.787},
url = {https://doi.org/10.1103/RevModPhys.80.787},
}

@article{HILLERY,
title = "{Distribution functions in physics: Fundamentals}",
author = "{M. Hillery}",
journal = {Phys. Rep.},
volume = {106},
number = {3},
pages = {121 - 167},
year = {1984},
month = {Apr},
publisher = {Elsevier},
doi = {10.1016/0370-1573(84)90160-1},
url = {https://doi.org/10.1016/0370-1573(84)90160-1},
}

@Book{BNDP,
title = "{Quantum Fields in Curved Spaces}",
author = "{N. D. Birrell and P. C. W. Davies}",
publisher = {Cambridge University Press, New York},
year = {1982},
}

@article{Davies_1975,
title = "{Scalar production in Schwarzschild and Rindler metrics}",
author = "{P. C. W. Davies}",
year = 1975,
month = {apr},
publisher = {{IOP} Publishing},
volume = {8},
number = {4},
pages = {609--616},
journal = {Journal of Physics A: Mathematical and General},
doi = {10.1088/0305-4470/8/4/022},
url = {https://doi.org/10.1088/0305-4470/8/4/022},
}

@book{FSA1,
title = "{Aspects of Quantum Field Theory in Curved Space-Time}",
author = "{S. A. Fulling}",
publisher = {Cambridge University Press},
year = {1989},
}

@article{FSA,
title = "{Nonuniqueness of canonical field quantization in Riemannian space-time}",
author = "{S. A. Fulling}",
journal = {Phys. Rev. D},
volume = {7},
issue = {10},
pages = {2850--2862},
numpages = {0},
year = {1973},
month = {May},
publisher = {American Physical Society},
doi = {10.1103/PhysRevD.7.2850},
url = {https://link.aps.org/doi/10.1103/PhysRevD.7.2850},
}

@article{HS1,
title = "{Black hole explosions?}",
author = "{S. W. Hawking}",
journal = {Nature},
volume = {248},
number = {6},
pages = {30--31},
year = {1974},
month = {Jul},
publisher = {Nature},
doi = {10.1038/248030a0},
url = {https://doi.org/10.1038/248030a0},
}

@article{U95,
title = "{Maintaining coherence in quantum computers}",
author = "{W. G. Unruh}",
journal = {Phys. Rev. A},
volume = {51},
issue = {2},
pages = {992--997},
numpages = {0},
year = {1995},
month = {Feb},
publisher = {American Physical Society},
doi = {10.1103/PhysRevA.51.992},
url = {https://doi.org/10.1103/PhysRevA.51.992},
}

@article{UWG,
title = "{Notes on black-hole evaporation}",
author = "{W. G. Unruh}",
journal = {Phys. Rev. D},
volume = {14},
issue = {4},
pages = {870--892},
numpages = {0},
year = {1976},
month = {Aug},
publisher = {American Physical Society},
doi = {10.1103/PhysRevD.14.870},
url = {https://link.aps.org/doi/10.1103/PhysRevD.14.870},
}

@article{ZWPR,
title = "{Resonance interaction energy between two accelerated identical atoms in a coaccelerated frame and the Unruh effect}",
author = "{W. Zhou, R. Passante and L. Rizzuto}",
journal = {Phys. Rev. D},
volume = {94},
issue = {10},
pages = {105025},
numpages = {9},
year = {2016},
month = {Nov},
publisher = {American Physical Society},
doi = {10.1103/PhysRevD.94.105025},
url = {https://link.aps.org/doi/10.1103/PhysRevD.94.105025},
}
\newpage

\begin{widetext}
\section{Supplemental material}

\noindent To compute the system's decoherence, we must calculate the time evolution of the density operator $\varrho(\tau)$. This can be achieved by considering the interaction Hamiltonian we previously derived:
\begin{align}
H_{int}(\Omega, \textbf{k}_{\bot}) = \frac{i\Lambda_{em}}{\sqrt{V}} \sigma_z \otimes \sum_{\Omega \textbf{k}_{\bot}, \lambda} \frac{1}{\sqrt{2 \Omega}} \big( g_{\Omega \textbf{k}_{\bot}, \lambda} \hat{b}_{\Omega \textbf{k}_{\bot}, \lambda}^\dagger - g_{\Omega \textbf{k}_{\bot}, \lambda}^\ast \hat{b}_{\Omega \textbf{k}_{\bot}, \lambda} \big). \nonumber
\end{align}
In order to determine the evolution operator, we utilize the interaction picture expressing the interaction Hamiltonian as:
\begin{align}
H_{I} = e^{-i\tau H_{f}} H_{emd} \ e^{i\tau H_{f}},
\label{EQ20}
\end{align}
where
\begin{align}
H_{f}  = \one \otimes \sum_{\Omega \textbf{k}_{\bot}, \lambda} \Omega \hat{b}_{\Omega \textbf{k}_{\bot}, \lambda}^{\dagger} \hat{b}_{\Omega \textbf{k}_{\bot}, \lambda},
\label{EQ21}
\end{align}
and
\begin{align}
H_{emd} = \frac{\varepsilon}{2} \sigma_{z} \otimes \one + \frac{i\Lambda_{em}}{\sqrt{V}} \sigma_{z} \otimes \sum_{\Omega \textbf{k}_{\bot}, \lambda} \frac{1}{\sqrt{2\Omega}} \big( g_{\Omega \textbf{k}_{\bot}, \lambda} \hat{b}_{\Omega \textbf{k}_{\bot}, \lambda}^{\dagger} - g_{\Omega \textbf{k}_{\bot}, \lambda}^{\ast} \hat{b}_{\Omega \textbf{k}_{\bot}, \lambda} \big).
\label{EQ22}
\end{align}
Then, in the interaction picture, the evolution operator for the system,
\begin{align}
U(\tau) =\hat{T} \exp \Big( -i\int_{0}^{\tau} dt' H_{I}(t') \Big),
\label{EQ23}
\end{align}
can be written as \cite{AINGPBMDAXW_DADOUDWF_2020}:
\begin{align}
U(\tau) = \exp{\Big( -i\nu(\tau) + \frac{i\varepsilon}{2}\sigma_{z} \otimes \one + \sigma_{z} \otimes \sum_{\Omega \textbf{k}_{\bot}, \lambda} \big( \hat{b}_{\Omega \textbf{k}_{\bot}, \lambda}^{\dagger} \xi_{\Omega \textbf{k}_{\bot}, \lambda}(\tau) - \hat{b}_{\Omega \textbf{k}_{\bot}, \lambda} \xi_{\Omega \textbf{k}_{\bot}, \lambda}^{\ast}(\tau) \big) \Big)},
\label{EQ24}
\end{align}	
where
\begin{align}
\nu(\tau) =\frac{i}{2} \sum_{\Omega \textbf{k}_{\bot}, \lambda} \int_{0}^{\tau} \big( \xi_{\Omega \textbf{k}_{\bot}, \lambda}^{\ast}(t) \dot{\xi}_{\Omega \textbf{k}_{\bot}, \lambda}(t) - \xi_{\Omega \textbf{k}_{\bot}, \lambda}(t) \dot{\xi}_{\Omega \textbf{k}_{\bot}, \lambda}^{\ast}(t) \big) dt,
\label{EQ25}
\end{align}
and
\begin{align}
\xi_{\Omega \textbf{k}_{\bot}, \lambda}(\tau) = \ i\Lambda_{em} g_{\Omega \textbf{k}_{\bot}, \lambda} \frac{1 - e^{-i\Omega \tau}}{\Omega^{3/2} \sqrt{2V}}.
\label{EQ26}
\end{align}
Then, the entire system is governed by the Hamiltonian
\begin{align}
H_{Tot} = \frac{\varepsilon}{2} \sigma_{z} \otimes \one + \one \otimes \sum_{\Omega \mathbf{k}_{\bot}, \lambda} \Omega \hat{b}_{\Omega \textbf{k}_{\bot}, \lambda}^{\dagger} \hat{b}_{\Omega \textbf{k}_{\bot}, \lambda} + \frac{i \Lambda_{em}}{\sqrt{V}} \sigma_{z} \otimes \sum_{\Omega \textbf{k}_{\bot}, \lambda} \frac{1}{\sqrt{2 \Omega}}(g_{\Omega \textbf{k}_{\bot}, \lambda} \hat{b}_{\Omega \textbf{k}_{\bot}, \lambda}^{\dagger} - g_{\Omega \textbf{k}_{\bot}, \lambda}^{\ast} \hat{b}_{\Omega \textbf{k}_{\bot}, \lambda}).
\label{EQ27}
\end{align}
Therefore, we assume that for the detector-electromagnetic-field system, the density operator $\varrho(\tau)$, at an initial time $\tau = \tau_0$, takes a form analogous to the scalar field case \cite{AINGPBMDAXW_DADOUDWF_2020}: $\varrho(\tau_{0}) = |\Psi_{0}\rangle \langle\Psi_{0}|$, with $|\Psi_{0}\rangle = |\psi_{0}\rangle \otimes |0_{M}\rangle$. Here, $|v_{0}\rangle = c_{1}|\downarrow\rangle + c_{2}|\uparrow\rangle$, represents the initial superpositional state of the detector and $|0_{M}\rangle$ denotes the Minkowski vacuum.
The density operator $\varrho(\tau)$ is also the detector's reduced density matrix, obtained by tracing out all vector field degrees of freedom. The time evolution of the matrix elements of the reduced density matrix is expressed as:
\begin{align}
\label{EQ28}
\rho_{ij}(\tau) = \langle i| {\rm Tr}_R U(\tau,\tau_0)\varrho(\tau_0) U^{-1}(\tau,\tau_0)|j \rangle, \ \quad \ (i,j=0,1),
\end{align}
where the index $i=0$ is associated with the eigenvector $|\downarrow\rangle$ and the index $i=1$ is associated with the eigenvector $|\uparrow\rangle$ of the operator $\sigma_{z}$.
The result of the action of  the evolution operator on an arbitrary pure initial state of the entire system can be obtained by a similar procedure developed in \cite{AINGPBMDAXW_DADOUDWF_2020}, and the resulting expressions are:
\begin{align}
\label{EQ29}
U(\tau) |\downarrow\rangle \otimes |\psi_{F}\rangle = & \ e^{-i\nu(\tau) -  i \varepsilon\tau/2} |\downarrow\rangle\otimes \prod_{\mathbf{k}} D \big( -\xi_{\Omega \textbf{k}_{\bot}, \lambda}(\tau)\big) |\psi_{F}\rangle, \\
U(\tau) |\uparrow\rangle \otimes |\psi_{F}\rangle = & \ e^{-i\nu(\tau) + i \varepsilon\tau/2} |\uparrow\rangle \otimes \prod_{\mathbf{k}} D \big( \xi_{\Omega \textbf{k}_{\bot}, \lambda}(\tau) \big) |\psi_{F}\rangle.
\label{EQ30}
\end{align}
Here, $|\psi_{F}\rangle$ is an initial state of the electromagnetic field and $D(\xi_{\Omega \textbf{k}_{\bot}, \lambda})$ denotes the displacement operator (defined for the scalar case in \cite{HILLERY}):
\begin{align}
D \big( \xi_{\Omega \textbf{k}_{\bot}, \lambda}(\tau) \big) = \exp \Big( \sum_{\lambda} \big( \xi_{\Omega \textbf{k}_{\bot}, \lambda}(\tau) \hat{b}_{\Omega \textbf{k}_{\bot}, \lambda}^{\dagger} - \xi_{\Omega \textbf{k}_{\bot}, \lambda}^{\ast}(\tau) \hat{b}_{\Omega \textbf{k}_{\bot}, \lambda} \big) \Big).
\label{EQ31}
\end{align}
The interaction of the detector with the vector field does not excite the detector, and the detection of the Unruh effect is reduced to the study of the phase decoherence:
\begin{align}
\rho_{01}(\tau) = e^{i \epsilon \tau - \gamma (\tau)} \rho_{01}(\tau_0).
\label{EQ32}
\end{align}
It is easy to prove that (\cite{AINGPBMDAXW_DADOUDWF_2020})
\begin{align}
\hat{b}_{\Omega \textbf{k}_{\bot}, \lambda} = \frac{\hat{a}_{-\omega \textbf{k}, \lambda} + e^{-\pi\Omega/a} \hat{a}_{\omega \mathbf{k}, \lambda}^{\dagger}}{\sqrt{1 - e^{-2\pi\Omega/a}}}.
\label{EQ33}
\end{align}
We also have calculated the Bogolyubov transformations and the mean number of particles in the mode $\Omega$:
\begin{align}
\tilde{n}_{\Omega} = \langle{0_M}| \hat{b}_{\Omega \textbf{k}_{\bot}, \lambda}^\dagger \hat{b}_{\Omega \textbf{k}_{\bot}, \lambda} |{0_M}\rangle = \sum_{\omega} |\beta_{\Omega \omega}|^{2} = \frac{1}{e^{2 \pi \Omega / a} - 1}.
\label{EQ34}
\end{align}
We can also easily prove that \cite{CLH}
\begin{align}
\alpha_{\Omega \omega \textbf{k}} = -e^{\pi \Omega / a} \beta_{\Omega \omega \textbf{k}},
\label{EQ35}
\end{align}
and
\begin{align}
\sum_{\omega} \alpha_{\Omega \omega \textbf{k}} \beta_{\Omega \omega \textbf{k}} = \sum_{\omega} \alpha_{\Omega \omega \textbf{k}}^{\ast} \beta_{\Omega \omega \textbf{k}}^{\ast} = 0.
\label{EQ36}
\end{align}
Leveraging the derived Bogolyubov relations, we obtain the decoherence function for a point-like detector, $\gamma_{0}(\tau)$, after some algebra, as follows:
\begin{align}
\gamma_{0}(\tau) = 2 \sum_{\Omega \textbf{k}_{\bot}} |\sum_{\lambda} \xi_{\Omega \textbf{k}_{\bot}, \lambda}(\tau)|^2 (1 + 2\tilde{n}_{\Omega}) = \sum_{\Omega \textbf{k}_{\bot}} \frac{\Lambda_{em}}{\Omega^{3} V} |\sum_{\lambda} g_{\Omega \textbf{k}_{\bot}, \lambda}|^2 \big( 1 - \cos{(\Omega \tau)} \big) \coth \Big( \frac{\pi \Omega}{a} \Big).
\label{EQ37}
\end{align}
Passing to the continuous limit, the sum over $(\Omega, \textbf{k}_{\bot})$ is replaced by an integral,
\begin{align}
\sum \rightarrow V/(2\pi)^3 \int d\Omega d^2\textbf{k}_{\bot}, \nonumber
\end{align}
and writing down the definition of the axial component $g_{\Omega}$ for our uniformly accelerated observer as
\begin{align}
& |g_{\Omega}|^2 = \int \Big|\sum_{\lambda} g_{\Omega \textbf{k}_{\bot}, \lambda}\Big|^2 d^2 {\mathbf{k}}_{\perp} = \int \Big|\sum_{\lambda} g_{\Omega \textbf{k}_{\bot}} \big[ \hat{\mathbf{k}} \times \varepsilon_{\Omega \textbf{k}_{\bot}, \lambda} \big]_{\zeta}\Big|^2 d^2 {\mathbf{k}}_{\perp},
\label{EQ38}
\end{align}
the decoherence function is then given by the following integral:
\begin{align}
\gamma_{0}(\tau) = \frac{\Lambda_{em}^{2}}{4 \pi^3} \int_0^\infty \frac{\abs{g_{\Omega}}^2}{\Omega^{3}} \big( 1 - \cos{(\Omega \tau)} \big) \coth \Big( \frac{\pi \Omega}{a} \Big) d\Omega.
\label{EQ39}
\end{align}
\end{widetext}

\end{document}